\documentclass[useAMS,usenatbib]{mn2e}
\usepackage{txfonts}
\usepackage{graphicx}
\usepackage{subfigure}
\usepackage{multirow}

\newcommand\nodata{ ~$\cdots$~ }%

\begin{document}

\title[Discovery of Five Very Low Mass Binaries]{Discovery of Five Very Low Mass Close Binaries, Resolved in the Visible with Lucky Imaging\thanks{Based on observations made with the Nordic Optical Telescope, operated on the island of La Palma jointly by Denmark, Finland, Iceland, Norway, and Sweden, in the Spanish Observatorio del Roque de los Muchachos of the Instituto de Astrofisica de Canarias.}
}

\author[N.M. Law et al.]{N.M. Law, S.T. Hodgkin, C.D. Mackay \\ Institute of Astronomy, Cambridge, UK}
\date{Received - / Accepted -}

\maketitle

\begin{abstract}
We survey a sample of 32 M5-M8 stars with distance $<$ 40pc for companions with separations between 0.1'' and 1.5'' and with $\rm{\Delta m_i<5}$. We find five new binaries with separations between 0.15'' and 1.1'', including a candidate brown dwarf companion. The raw binary fraction is $16^{+8}_{-4}\%$ and the distance bias corrected fraction is $7^{+7}_{-3}\%$, for companions within the surveyed range. No systems with contrast ratio $\rm{\Delta m_i>1}$ were found, even though our survey is sensitive to $\rm{\Delta m\leq5}$ (well into the brown dwarf regime). The distribution of orbital radii is in broad agreement with previous results, with most systems at 1-5AU, but one detected binary is very wide at $46.8\pm5.0$AU. We also serendipitously imaged for the first time a companion to Ross 530, a metal-poor single-lined spectroscopic binary. We used the new Lucky Imaging system LuckyCam on the 2.5m Nordic Optical Telescope to complete the 32 very low mass star SDSS i' and z' survey in only 5 hours of telescope time. 

\end{abstract}


\maketitle

\section{Introduction}

There are compelling reasons to search for companions to nearby stars. In particular, the properties of binary systems provide important clues to their formation processes. Any successful model of star formation must be able to account for both the frequency of multiple star systems and their properties (separation, eccentricity and so forth) -- as well as variations in those properties as a function of system mass. In addition, the orbits of binary systems provide us with the means to directly measure the mass of each component in the system. This is fundamental to the calibration of the mass-luminosity relation (MLR: \citealt{Henry_1993, Henry_1999, Segransan_2000}).

The stellar multiplicity fraction appears to decrease with decreasing primary mass (eg. \citealt{Siegler_2005}). Around 57\% of solar-type stars (F7--G9) have known stellar companions \citep{Abt_1976, Duquennoy_1991}, while imaging and radial velocity surveys of early M dwarfs suggest that between 25\% \& 42\% have companions \citep{Henry_1990, Fischer_1992, Leinert_1997, Reid_1997}. Later spectral types have been studied primarily with high resolution adaptive optics imaging: \citealt{Close_2003} and \citealt{Siegler_2005} find binary fractions of around $10$--$20\%$ for primary spectral types in the range M6--L1. \citealt{Bouy_2003} and \citealt{Gizis_2003} find that 10--15\% of L dwarfs have companions, and \citealt{Burgasser_2003} find that 10\% of T dwarfs have binaries. These very low mass (VLM) \mbox{M, L and T} systems appear to have a tighter and closer distribution of orbital separations, peaking at around 4\,AU compared to 30\,AU for G dwarfs \citep{Close_2003}. 

However, each of these surveys have inevitably different (and hard to quantify) sensitivities, the effect of which is especially evident in the large spread in the derived multiplicity of early M-dwarfs. In particular, high-resolution imaging surveys are sensitive only to companions wider than $\sim$0.1'' while radial velocity surveys are much more sensitive to closer (shorter period) companions. Maxted \& Jeffries\,(2005), by examining a small sample of radial velocity measurements, estimate that accounting for systems with $a<3$\,AU could increase the overall observed VLM star/BD binary frequency to 32--45\%. In addition, each survey's target sample has a different selection of target stellar parameters (as well as different incompletenesses and biases), leading to difficulties in the comparison and pooling of results for the surveys.

This paper details the first results of an ongoing effort by our group to greatly increase the number of known VLM binary systems. A large sample size is made possible by the uniquely low observation overheads offered by the new high-resolution imaging system LuckyCam. We present results here from a trial 32-star sample, completed in only 5 hours of on-sky time. The detected binaries have been very briefly described as part of a larger sample in \citet{Law_binaries_05}; we here undertake detailed investigations of the systems and sample.

\begin{figure}
  \centering
  \resizebox{\columnwidth}{!}
   {
	\includegraphics{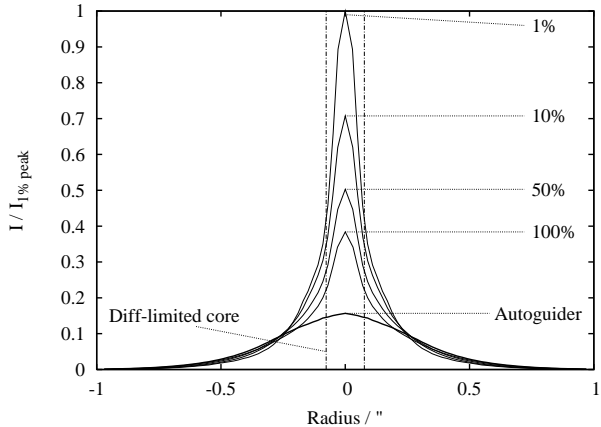}
   }
   \caption{Typical radial Lucky Imaging profiles, azimuthally averaged from the point spread function of a star imaged at 30Hz in 0.6'' seeing. 100\% selection corresponds to fast shift-and-add (tip/tilt correction) imaging. When selecting 1\% of frames the Strehl ratio is almost tripled relative to the 100\% selection, while the light from the star is concentrated into an area approximately four times smaller.}
   \label{FIG:Profiles}
\end{figure}

Our Lucky Imaging system, LuckyCam, takes a sequence of images at $>$10 frames per second using a very low noise L3CCD based conventional camera. Because the atmospheric turbulence affecting the images changes on very short timescales, there are rapid ($<$100ms coherence time) variations in the image quality of the frames. To construct a high resolution long exposure image we select, align and co-add only those frames which meet a quality criterion. By varying the criterion we can trade off sensitivity against ultimate resolution. The technique is described in more detail in \citet{Tubbs_2002, Law05}. Lucky Imaging is an entirely passive technique, allowing data to taken as soon as the telescope is pointed, and thus leading to very low time overheads. With stars brighter than $\rm{+15m}$ in $<0.6''$ seeing, resolutions very close to the diffraction limit (Strehl ratios $>$ 0.1) of the 2.5m Nordic Optical Telescope (NOT) in i' band are regularly obtained. Figure \ref{FIG:Profiles} shows examples of the general form of the Lucky Imaging point spread function (PSF). 

In this paper we present five new VLM binaries and evaluate the utility of LuckyCam for programmes of this type. In section \ref{SEC:Sample} we define the sample of VLM stars imaged in this survey. In section \ref{SEC:Obs} we describe the observations, Lucky Imaging, the data reduction techniques and the survey sensitivity. Section \ref{SEC:Results} describes the results of our survey and details the properties of the detected binaries. In section \ref{SEC:Discussion} we discuss the results in the context of other surveys. We conclude in section \ref{SEC:Concs}.
\begin{figure}
  \centering
  \resizebox{\columnwidth}{!}
   {
	\includegraphics[angle=-90]{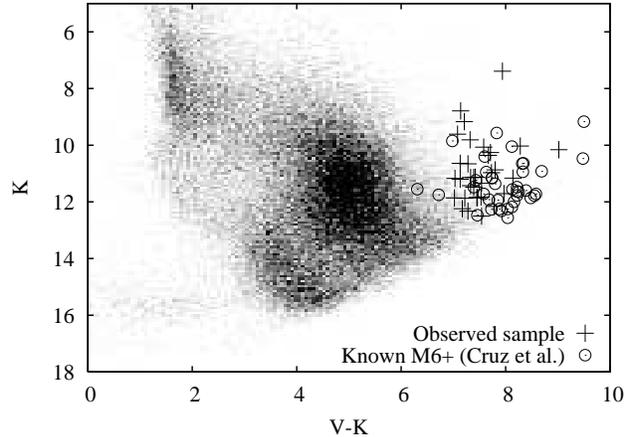}
   }
   \caption{The observed sample (crosses), plotted in a V/V-K colour-magnitude diagram. The background distribution shows all stars in the LSPM-North catalogue. Circles show 38 spectroscopically confirmed M6 and later dwarfs from \citealt{Cruz03}, which also appear in the LSPM-North; all but two (both M6) are recovered by our selection criteria.}
   \label{FIG:Sample}
\end{figure}

\begin{table*}
 \centering
  \caption{The observed sample. The quoted V \& K magnitudes are taken from the LSPM catalogue. K magnitudes are based on 2MASS photometry; the LSPM-North V-band photometry is estimated from photographic $\rm{B_J}$ and $\rm{R_F}$ magnitudes and is thus approximate only, but is sufficient for spectral type estimation (figure \ref{FIG:Sample}). Spectral types and distances are estimated from the V \& K photometry and the young-disk photometric parallax relations described in \citet{Leggett_1992}. Spectral types are accurate to approximately 0.5 spectral classes and distances to $\sim30$\%.  \label{Tab:Sample}}
  \begin{tabular}{llrrllrc}
  \hline
   LSPM ID & 2MASS ID & V & V-K & PM / arcsec/yr& Estimated spectral type & Photom. dist/pc & Newly detected companion? \\
  \hline
LSPM J1235+1318 & 12351726+131805 &  18.0 & 7.7 & 0.219 & M6.5 & 14 & $\ast$ \\ 
LSPM J1235+1709 & 12351850+170937 &  19.3 & 7.5 & 0.570 & M6.5 & 29 &      \\
LSPM J1246+0706 & 12460939+070624 &  17.7 & 7.1 & 0.549 & M6.0 & 19 &    \\
LSPM J1303+2414 & 13034100+241402 &  19.6 & 7.9 & 0.370 & M7.0 & 24 &     \\
LSPM J1305+1934 & 13053667+193456 &  18.4 & 7.3 & 0.554 & M6.5 & 22 &     \\
LSPM J1314+1320 & 13142039+132001 &  15.9 & 7.2 & 0.307 & M6.0 & 7.7 & $\ast$ \\
LSPM J1336+1022 & 13365393+102251 &  18.7 & 7.3 & 0.381 & M6.5 & 26 &     \\
LSPM J1341+0805 & 13413291+080504 &  18.5 & 7.4 & 0.269 & M6.5 & 22 &     \\ 
LSPM J1354+0846 & 13540876+084608 &  19.3 & 8.2 & 0.219 & M7.0 & 17 &     \\ 
LSPM J1423+1318 & 14231683+131809 &  17.9 & 7.3 & 0.174 & M6.5 & 18 & $\ast$ \\ 
LSPM J1423+1426 & 14234378+142651 &  17.9 & 7.7 & 0.638 & M6.5 & 13 &     \\
LSPM J1428+1356 & 14280419+135613 &  18.3 & 8.3 & 0.605 & M8.0 & 10 &     \\ 
LSPM J1432+0811 & 14320849+081131 &  16.3 & 7.2 & 0.455 & M6.0 & 9.2 &     \\ 
LSPM J1440+1339 & 14402293+133923 &  19.0 & 7.7 & 0.337 & M6.5 & 22 &     \\ 
LSPM J1454+2852 & 14542356+285159 &  18.6 & 7.0 & 0.212 & M6.0 & 32 &      \\ 
LSPM J1516+3910 & 15164073+391048 &  17.1 & 7.3 & 0.224 & M6.5 & 12 &     \\ 
LSPM J1554+1639 & 15540031+163950 &  19.9 & 7.8 & 0.529 & M7.0 & 30 &     \\ 
LSPM J1605+6912 & 16050677+691232 &  19.3 & 7.5 & 0.224 & M6.5 & 29 &     \\ 
LSPM J1606+4054 & 16063390+405421 &  17.6 & 7.6 & 0.735 & M6.5 & 12 &     \\ 
LSPM J1622+4934 & 16225554+493457 &  19.4 & 7.2 & 0.316 & M6.0 & 39 &     \\ 
LSPM J1626+2512 & 16263531+251235 &  19.3 & 7.5 & 0.271 & M6.5 & 29 &     \\ 
LSPM J1646+3434 & 16463154+343455 &  16.6 & 7.0 & 0.550 & M6.0 & 13 &     \\ 
LSPM J1647+4117 & 16470576+411706 &  18.5 & 7.4 & 0.289 & M6.5 & 22 &     \\ 
LSPM J1653+0000 & 16531534+000014 &  18.6 & 7.8 & 0.287 & M7.0 & 17 &     \\ 
LSPM J1657+2448 & 16572919+244850 &  18.8 & 7.5 & 0.391 & M7.5 & 23 &     \\ 
LSPM J1703+5910 & 17031418+591048 &  18.8 & 7.0 & 0.572 & M6.0 & 35 &     \\ 
LSPM J1735+2634 & 17351296+263447 &  19.1 & 9.0 & 0.349 & M9.0 & 9.2 & $\ast$ \\
LSPM J1741+0940 & 17415439+094053 &  18.7 & 7.8 & 0.435 & M7.0 & 17 &    \\ 
LSPM J1758+3157 & 17580020+315726 &  18.2 & 7.1 & 0.158 & M6.0 & 24 &    \\ 
LSPM J1809+2128 & 18095137+212806 &  18.3 & 7.1 & 0.193 & M6.0 & 25 & $\ast$ \\ 
LSPM J1816+2118 & 18161901+211816 &  19.0 & 7.2 & 0.171 & M6.0 & 32 &      \\ 
LSPM J1845+3853 & 18451889+385324 &  19.4 & 8.4 & 0.408 & M8.0 & 16 &     \\ 

 \hline
\end{tabular}
\end{table*}

\section{The Sample}
\label{SEC:Sample}
We selected a distance, flux and colour limited sample of stars from the LSPM-North Catalogue \citep{Lepine05}, which is the result of a systematic search for stars with declination $>$ 0 and proper motion $>$ 0.15''/year in the Digitized Sky Surveys. Most stars in the catalogue have 2MASS IR photometry as well as V-band magnitudes estimated from the photographic $\rm{B_J}$ and $\rm{R_F}$ bands.

The properties of the selected stars are detailed below:

\begin{enumerate}
\item{$\rm{V-K > 7}$; thus selecting approximately M6 and later stars \citep{Leggett_1992}. The LSPM-North proper motion cut ensures that all stars are relatively nearby and thus removes giant stars from the sample.}
\item{$\rm{Distance < 40pc}$. Absolute magnitudes are estimated from the V-K colours quoted in the LSPM and the V-K vs. $\rm{M_K}$ relations described in \citet{Leggett_1992}. Distances are then estimated by comparing the estimated absolute magnitude to the observed K magnitude.}
\item{$\rm{m_i < +15.5}$; Lucky Imaging requires a $\rm{m_i = 15.5}$ guide star for full performance. All targets serve as their own guide star.}
\item{We removed all stars from the remaining sample that had been to our knowledge previously observed at high angular resolution.}

\end{enumerate}

\begin{figure*}
  \centering
	\subfigure{\resizebox{\columnwidth}{!}{\includegraphics{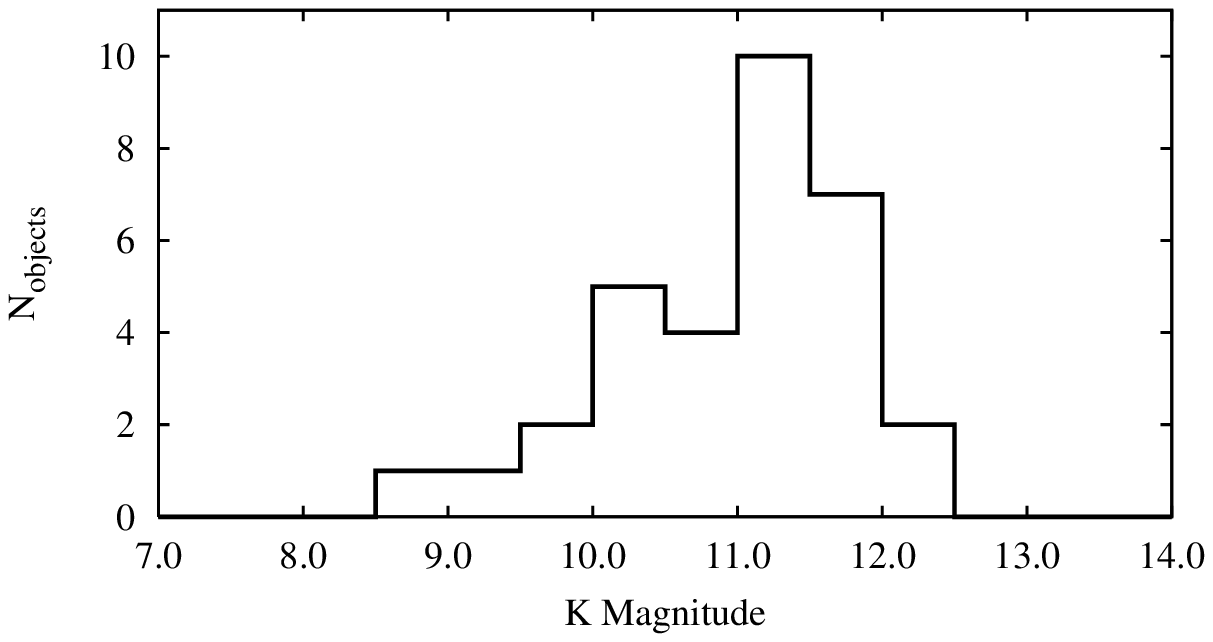}}}
	\subfigure{\resizebox{\columnwidth}{!}{\includegraphics{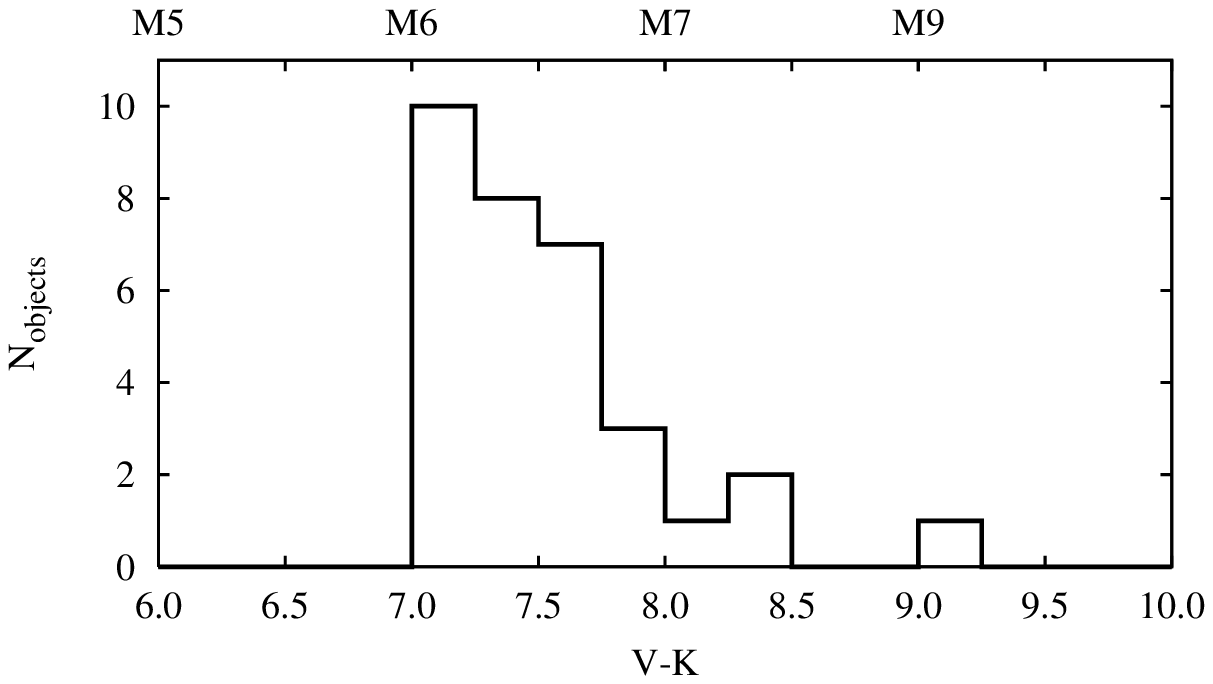}}}
   \caption{The distribution of the target sample in K magnitude and V-K colour, both from LSPM-North photometry.}
   \label{FIG:sample_hists}
\end{figure*}

The remaining sample consists of 91 stars in the R.A. and declination range that was accessible to us during the survey. 32 were selected for these observations, and are detailed in table \ref{Tab:Sample}. The region of colour-magnitude space in which they are found is shown in figure \ref{FIG:Sample}; the distributions of magnitudes and colours are detailed in figure \ref{FIG:sample_hists}.

We note that the V-band LSPM photometry has been estimated from observations in the photographic $\rm{B_J}$ and $\rm{R_F}$ bands (as detailed in \citealt{Lepine05}), and its use therefore requires some caution. To test its utility for late M-dwarf target selection we have confirmed that a sample of spectroscopically confirmed late M-dwarfs \citep{Cruz03} is fully recovered by our V-K selection (figure \ref{FIG:Sample}). In addition, LuckyCam resolved SDSS i' and z' photometry gives confirmation of estimated spectral type for the objects in our full survey. In all checked cases the spectral type and distance estimated from LSPM-North V-K photometry matches that derived from LuckyCam SDSS i' and z' photometry.

\section{Observations}
\label{SEC:Obs}
We performed observations with the Cambridge Lucky Imaging system, LuckyCam, on the 2.56m Nordic Optical Telescope in June 2005, during 5 hours of on-sky time spread over 4 nights. Each target was observed for 100 seconds in each of the SDSS i' and z' filters. SDSS standard stars (\citealt{Smith_2002}) were observed for photometric calibration; globular clusters and similar fields were imaged for astrometric calibration. The seeing measured by the Isaac Newton Group RoboDIMM at the observatory site varied between 0.5'' and 1.0'' during the observations, with a median of $\rm{\sim}$0.8''. 

Each target observation was completed in an average of 10 minutes including telescope pointing, 100 seconds of integration in each filter, the observation of one standard star for every three targets, and all other overheads.

\subsection{LuckyCam}

LuckyCam is a new imaging system designed for the new high angular resolution technique Lucky Imaging \citep{Law05, Tubbs_2002}. An electron-multiplying L3CCD is mounted at the focus of a simple reimaging camera. The effectively zero readout noise L3CCD allows guide stars as faint as i'=+15.5m to be used for high angular resolution observations.

For this survey we operated LuckyCam at 30 frames per second, with each frame being 552x360 pixels. The image scale was 0.04"/pixel, giving a field of view of 22$\times$14.4 $\rm{{arcseconds^2}}$. All frames were recorded to a fast RAID array and stored for later reduction; the observations totalled approximately 100GB. 

The dataset was reduced with the standard Lucky Imaging pipeline, described in detail in \citet{Law05}. Briefly, frames are flat fielded and bias is removed. The frames are then categorised by image quality; the highest quality frames are aligned and added using the Drizzle algorithm \citep{Drizzle} to give a final image sampled at 0.02''/pixel.

\subsection{Binary detection and photometry}
Obvious binaries with a low contrast ratio and/or $>$ 0.5'' separation were detected by eye in reduced images including $<=10\%$ of frames. We limit the companion detection radius to 1.5''. Using a small detection area allows us to use the remainder of our 20''x14'' fields as control areas; because the chance probability of an object falling within our small detection radius is very low, we can then show that any detections are likely to be physically associated with the target star.

In order to detect fainter companions we fit and subtract a model Moffat profile point spread function (e.g. \citealt{Law05}) to each target, using 50\% of the recorded frames to increase the SNR of faint companions at the expense of some resolution. Candidate companions were detected using a sliding-box method, with custom software implementing the detection criteria.

We stipulated the detection of a faint companion to require a $10\sigma$ deviation above the background noise, which is due to both photon and speckle noise and varies with distance from the primary star. The background noise at each radius was specified to be the upper $1\sigma$ excursion from the average RMS noise at several azimuthal positions.

In addition, the following criteria were implemented to confirm the detection of faint companions:
\begin{enumerate}
  \item{the candidate must maintain a constant flux per frame as more frames are included in the reduced images - most persistent speckles appear in only a fraction of frames and so fail this test.}
  \item{detection must be repeated in the same position in each filter.}
  \item{the candidate must appear point-like (extended objects are thus not detected).}
  \item{the candidate must not be visible in the PSFs of other stars observed within a few minutes of the target.}
\end{enumerate}

We acquired resolved flux measurements and errors for binaries wider than 0.4'' with simple aperture photometry. However, at closer radii more sophisticated strategies are required, especially since four of the five detected binaries have primaries fainter than i'=15m. If a Lucky Imaging guide star is faint, its PSF is altered by frame selection proceeding partially on the basis of high excursions of photon-shot-noise. Since the companion and primary now have different PSFs point spread function subtraction is difficult (although possible with sufficiently similar calibrator binary observations). The closest binary detected in these observations was sufficiently bright to avoid these problems, however.

Extensive experimentation confirmed that simple aperture photometry also provides accurate flux measurements at close radii for Lucky Imaging PSFs, provided that care is taken in the choice of foreground and background aperture sizes. The two close binaries with relatively faint primaries were reduced in this manner.

The accuracy and precision of the derived contrast ratios was measured by repetition over several different frame selection fractions (and thus several different PSFs). The accuracy of the algorithms was also checked against simulated binary images.

\begin{figure}
  \centering
  \resizebox{\columnwidth}{!}
   {
	\includegraphics{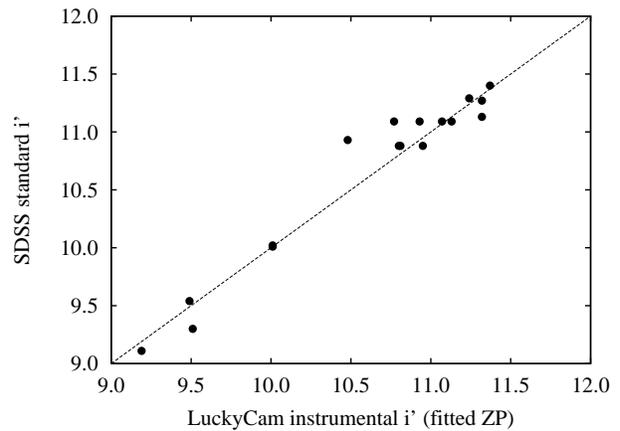}
   }
   \caption{Calibration of the LuckyCam photometry against SDSS standards from \citet{Smith_2002}. The dashed line has a gradient of 1.0. As all our targets are very red the standards with the highest i'-z' available were used; there is no i'-z' colour term detectable within the photometric errors. The photometric accuracy of the LuckyCam system has also been verified with much fainter targets.}
   \label{FIG:SDSS_stds}
\end{figure}

\begin{figure}
  \centering
  \resizebox{\columnwidth}{!}
   {
	\includegraphics{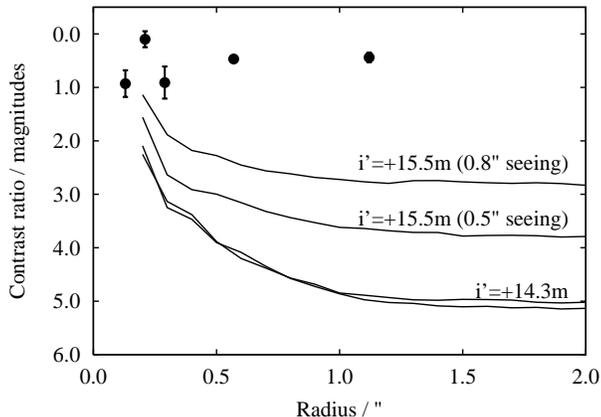}
   }
   \caption{The $\rm{10\sigma}$ contrast ratios achieved in a variety of conditions. The i'=+14.3m star is shown in both 0.5'' and 0.8'' seeing; the imaging performance is not dependent on this small difference in seeing for bright stars. However, poorer seeing reduces the average light per pixel sufficiently to affect the faint guide star imaging performance. The detected binaries are also shown (the measured contrast ratio uncertainties are often smaller than the plotted points). At radii $<$ 0.2'' the cell-size of the faint companion detection algorithm does not adequately sample the shape of the PSF; detectable contrast ratios in this area are approximately 2 magnitudes.}
   \label{FIG:Sens_curves}
\end{figure}

\begin{figure}
  \centering
  \resizebox{\columnwidth}{!}
   {
	\includegraphics{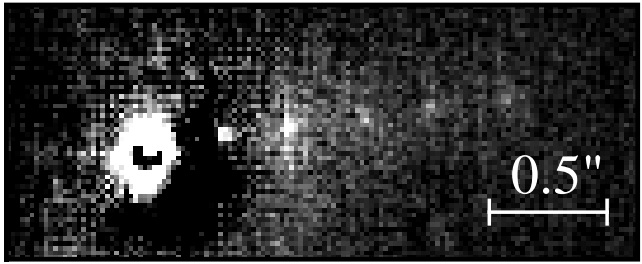}
   }
   \caption{Simulated faint companion PSFs around an i'=14m star. At each radius in each of the 2800 frames in this observation we added to the background a version of the primary PSF, rescaled to the $\rm{10\sigma}$ detection flux shown in figure \ref{FIG:Sens_curves} and with appropriate Poisson and L3CCD multiplication noise added. For clarity the primary's PSF has been subtracted. The simulated companions become brighter at lower radii to allow detection above both photon and speckle noise from the primary's PSF. The residual ring around the central star is the Airy ring expected in near diffraction limited images, which is not included in the model PSF.}
   \label{FIG:Sens_sims}
\end{figure}

Photometry in the SDSS system was calculated from the total integrated flux in a 3'' radius aperture, calibrated against SDSS standards (\citealt{Smith_2002}, figure \ref{FIG:SDSS_stds}). We then use the measured contrast ratios to derive resolved photometry. In each observation the L3CCD gain is calibrated by fitting a theoretical model to the histogram of single-photon data values (for details on the L3CCD statistics see eg. \citealt{Basden_PC}). The calibrated magnitudes are then calculated as:
\begin{equation}
\rm{mag = ZP - 2.5log_{10}(gain \times DN/sec)}
\end{equation}
where the raw photometry is given in data numbers (DN) per second, gain is measured in photons per DN, and ZP is the photometric zero point of the system. Airmass corrections are not calculated because all observations were performed within $\rm{30^o}$ of the zenith; the approximately 10\% uncertainty in the L3CCD gain calibration dominated the remaining errors.

\subsection{Sensitivity}
Beyond 1.0'' radius from the primary detection sensitivity is primarily limited by the sky background; at smaller radii both azimuthal variations in the target star's PSF and its photon shot noise limit the detection sensitivity. The SDSS i' detection contrast ratios for two typical stars are shown in figure \ref{FIG:Sens_curves} and example faint companion simulated PSFs are shown in figure \ref{FIG:Sens_sims}. 

Around a star with $\rm{m_i} \leq 14$, the survey is sensitive to $\Delta m_i = 5$ at radii $>$ 1.0'' and $\Delta m_i = 4$ at $>$ 0.5'', corresponding to the detection of brown dwarf companions around all the surveyed stars. For example, at an age of 5.0Gyr an object at the hydrogen-burning limit has $\rm{M_i \approx 16}$. Typical stars in this sample have $M_i \approx 13-15$. Figure \ref{FIG:Sens_curves} thus implies the survey is sensitive to brown dwarf companions at $>$ 1.0'' around all surveyed stars -- and much closer (or, equivalently, much fainter) brown dwarf companions around many of the stars. 

\section{Results \& Analysis}
\label{SEC:Results}
\begin{figure*}
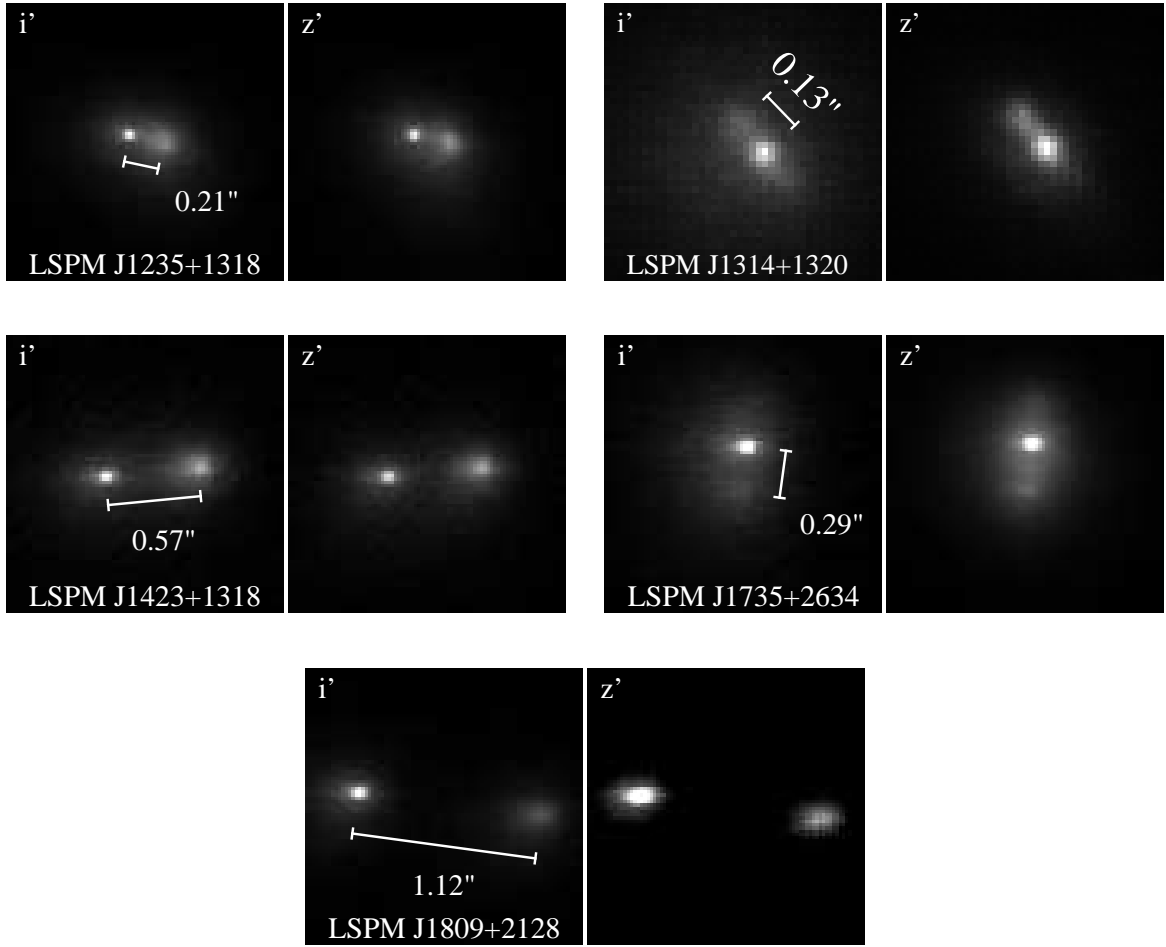

  \centering
	\subfigure{\resizebox{1.45in}{!}{\includegraphics{R57i.eps}}}
	\subfigure{\resizebox{1.45in}{!}{\includegraphics{R57z.eps}}}
	\hspace{0.15in}
	\subfigure{\resizebox{1.45in}{!}{\includegraphics{R69i.eps}}}
	\subfigure{\resizebox{1.45in}{!}{\includegraphics{R69z.eps}}}
	\subfigure{\resizebox{1.45in}{!}{\includegraphics{R86i.eps}}}
	\subfigure{\resizebox{1.45in}{!}{\includegraphics{R86z.eps}}}
	\hspace{0.15in}
	\subfigure{\resizebox{1.45in}{!}{\includegraphics{R124i.eps}}}
	\subfigure{\resizebox{1.45in}{!}{\includegraphics{R124z.eps}}}
	\subfigure{\resizebox{1.45in}{!}{\includegraphics{R133i.eps}}}
	\subfigure{\resizebox{1.45in}{!}{\includegraphics{R133z.eps}}}
   \caption{Lucky imaging of the five new binaries. All images are a 10\% selection from 3000 frames taken at 30FPS, with the exception of the 0.13'' binary LSPM J1314+1320 which required a more stringent selection. The very close binary's i' image is the result of a 2.5\% selection from 3000 frames and its z' image is a 1.0\% selection from 10,000 frames taken at the higher frame rate of 50FPS. In each case the image greyscale covers the full dynamic range of the images. All images are orientated with North up and East left. In many cases the guide star has a very compact core, due to the Lucky Imaging system aligning images partially on the basis of high photon-shot-noise excursions.}
   \label{FIG:binary_images}
\end{figure*}

We detected five previously unknown binaries in our VLM star sample, with separations ranging from 0.13'' to 1.12''. Resolved i' and z' images of each binary are given in figure \ref{FIG:binary_images}; their directly observed properties are summarised in table \ref{Tab:lucky_binary_obs}. 
\subsection{Confirmation of the binaries}
In our entire observation set (including some fields from a different sample not presented in this paper), covering $\rm{(22''\times14.4'')\times48\,fields}$, we detect only one field object which is sufficiently red to be characterised as a companion, should it have fallen close to a target star (the object is in the sample not presented here). Limiting our detection radius to 1.5'', and thus reducing the likelihood of chance associations, the probability of one or more false associations in our dataset (i.e. the probability of any number of our binaries not being physically associated) is therefore only 1.5\%.

All detected companions have PSFs consistent with being unresolved point sources, although in many cases the (faint) guide star has a more compact core, due to the Lucky Imaging system aligning images partially on the basis of high photon-shot-noise excursions.

We also find below that the photometric parallaxes of each binary's components are equal (within the stated errors), and they are therefore at equal distances, further decreasing the likelihood of contamination. Thus, we conclude that all the candidate binaries are physically associated systems. As all of these systems have high proper motion, confirmation of common proper motion can be easily determined from repeat measurements on year timescales.

\subsection{Distances, Ages and Masses}
We show derived properties of the individual binary components in table \ref{Tab:lucky_component_properties}. Masses of each component are estimated by comparing the component absolute magnitudes with the models presented in \citet{Baraffe98}, custom-integrated over the SDSS i' passband (I. Baraffe, private communication). In the absence of published age estimates for these targets the full range of ages found in the solar neighbourhood (with 5 Gyr as our adopted best-estimate value) is assumed (figure \ref{FIG:Mass_model}).

None of the 5 systems have trigonometric parallaxes, so in table \ref{Tab:lucky_component_properties} we estimate distances using the i'-z' vs. $\rm{M_i}$ relations given in \citet{Hawley_2002}. In all cases the binary components were found to be at equal distances, within the stated errors, and we combine their values for a single more precise system distance.

\subsection{Notes on the new systems}
None of these stars have been previously investigated in detail in published work; no previous high resolution imaging or spectroscopy (resolved or unresolved) is available. 

\noindent\textbf{LSPM J1235+1318}
A close to equal-mass 0.21'' binary with an estimated distance of $24.4\pm2.5$pc and an estimated orbital radius of $5.1\pm0.9$AU.

\noindent\textbf{LSPM J1314+1320}
Resolved at a separation of only 0.13'' and is one of the two nearest systems presented here ($9.8\pm2.0$pc), with an orbital radius of only $1.3\pm0.3$AU. SDSS i' photometry could not be calculated because (in this observation only) the camera gain was set too low to accurately calibrate. This system's distance has been estimated from the V \& K magnitudes listed in the LSPM North and the young disk photometric distance relations from \citet{Leggett_1992}, assuming that the two components have equal V-K colours.  

\noindent\textbf{LSPM J1423+1318}
A 0.57'' binary estimated to be M5.5/M5.5 with an estimated distance $33.1\pm3.4$pc and an orbital radius of $18.9\pm2.0$AU. 

\begin{figure}
  \centering
  \resizebox{\columnwidth}{!}
   {
	\includegraphics{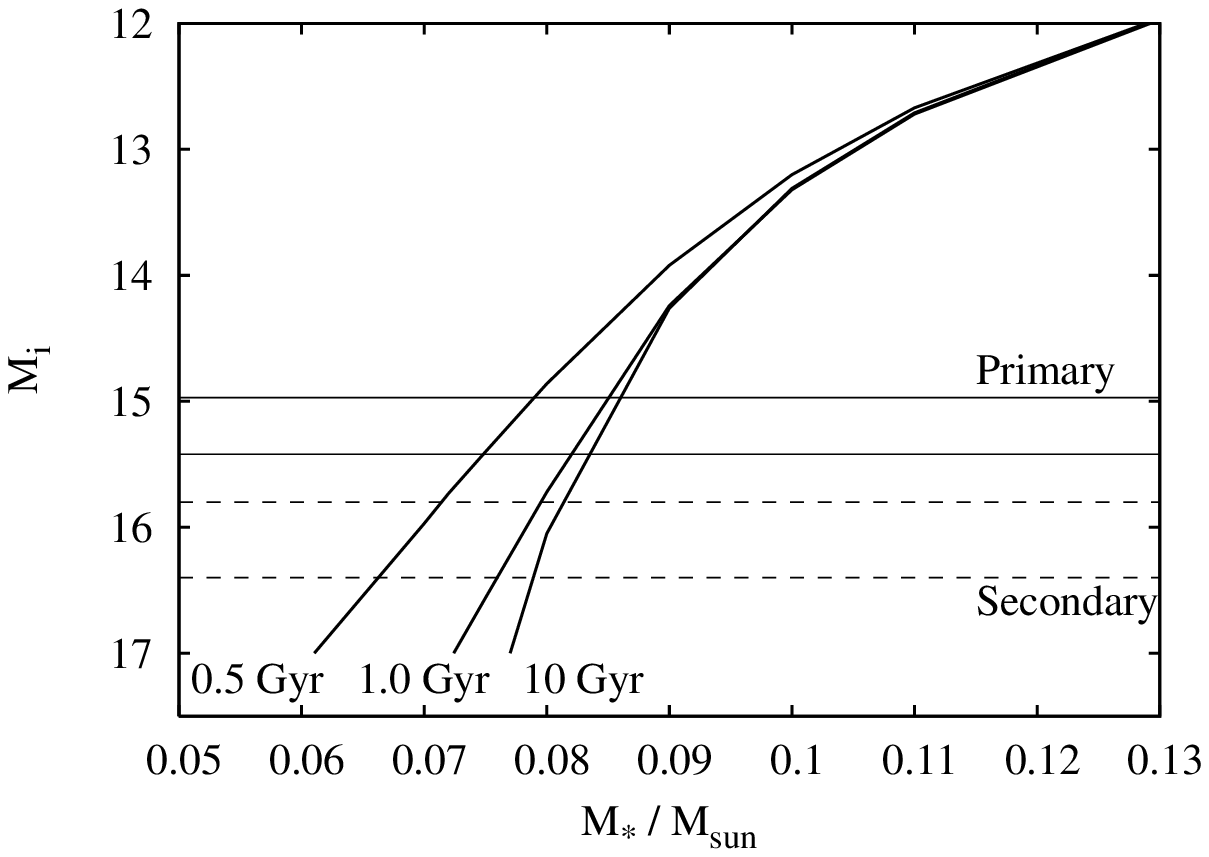}
   }
   \caption{The allowed mass ranges for the components of LSPM J1735+2634. The solid horizontal lines delimit the allowed mass range for the system's primary, given its calculated absolute magnitude; the dashed lines show the mass range of the secondary. The isochrones are from \citet{Baraffe98}; we here assume an age range covering the distribution in the solar neighbourhood, 0.5-10.0 Gyr \citep{G2002izis}.}
   \label{FIG:Mass_model}
\end{figure}

\begin{table*}
 \centering
  \caption{Observations of the new binary systems. Errors are 1-sigma and derived from the variation in fit values with different frame selection fractions (and thus PSFs). The position angle (P.A.) error includes a $\rm{1.0^o}$ uncertainty in the orientation calibration. The companion to LSPM J1735+2634 is very faint and so only upper and lower limits to the i' and z' contrast ratios are given. \label{Tab:lucky_binary_obs}}
  \begin{tabular}{lllllllr}

  \hline
  LSPM ID & 2MASS ID & $\rm{\Delta i'}$ & $\rm{\Delta z'}$ & Separation (arcsec) &  P.A. (deg) \\
  \hline
  LSPM J1235+1318 & 12351726+131805 & $0.10\pm0.15$ & $0.07\pm0.15$ & $0.21\pm0.03$ & $257.0\pm2.5$ \\ 
  LSPM J1314+1320 & 13142039+132001 & $0.93\pm0.25$ & $0.97\pm0.25$ & $0.13\pm0.02$ & $46.0\pm2.0$ \\ 
  LSPM J1423+1318 & 14231683+131809 & $0.47\pm0.03$ & $0.48\pm0.05$ & $0.57\pm0.01$ & $275.6\pm2.0$ \\
  LSPM J1735+2634 & 17351296+263447 & $0.61-1.21$ & $0.62-1.11$ & $0.29\pm0.01$ &     $171.2\pm2.1$ \\ 
  LSPM J1809+2128 & 18095137+212806 & $0.44\pm0.09$ & $0.40\pm0.10$ & $1.12\pm0.01$ & $262.7\pm2.0$ \\
 \hline
\end{tabular}
\end{table*}

\label{SEC:Analysis}

\begin{table*}
 \centering
  \caption{The new binary systems' component properties. Spectral types and distances have been derived from the i'-z' relations in \citet{Hawley_2002}. The spectral type relations have a plateau in i'-z' at L0-L3, limiting precision for the later spectral classes. We further constrained the earliest spectral types (M4-M6) by comparing the derived i' and z' absolute magnitudes to those expected for those spectral types \citep{Hawley_2002}. Orbital radius uncertainties are $\rm{1\sigma}$ and are estimated in quadrature from the system distance and separation uncertainties. Masses are estimated from the allowed ranges given by the models of \citealt{Baraffe98} (figure \ref{FIG:Mass_model}), between the $1\sigma$ photometric errors and the range of ages in the solar neighbourhood. Values of $q$ $>$ 1.0 imply that the true primary of the system may have been identified as the secondary. LSPM J1735+2634B is very faint, leading to difficulties in estimating resolved photometry, and so we only note upper and lower limits to the system's observed and derived properties. The L3CCD gain for the i' observation of LSPM J1314+1320 was set too low to accurately calibrate, so we do not give i' photometry for this system.\label{Tab:lucky_component_properties}
}
  \begin{tabular}{lllllllll}
  \hline
  LSPM ID & SDSS i' mag & SDSS z' mag & i' - z' & Spectral type & Mass / $\rm{M_{\odot}}$ & $q$ ($\rm{M_s / M_p}$) & Distance / pc & orbital radius / AU  \\
  \hline
  LSPM J1235+1318A & $15.1\pm0.20$ & $13.8\pm0.20$ & $1.25\pm0.23$ & $\rm{M5-M7}$ & $0.097-0.107$ & $0.89-1.08$ & $24.4\pm2.7$ & $5.1\pm0.9$ \vspace{2pt} \\ 
  LSPM J1235+1318B & $15.2\pm0.13$ & $13.9\pm0.13$ & $1.30\pm0.18$ & $\rm{M6-M7}$ & $0.097-0.106$ & & &   \\ 
\\
  LSPM J1314+1320A & \nodata & $11.9\pm0.32$ & \nodata & \nodata & \nodata & \nodata & $9.8\pm2.0$ & $1.3\pm0.3$ \vspace{2pt}\\ 
  LSPM J1314+1320B & \nodata & $12.8\pm0.21$ & \nodata & \nodata & \nodata & &   \\ 
\\
  LSPM J1423+1318A & $15.1\pm0.11$ & $13.9\pm0.12$ & $1.21\pm0.16$ & $\rm{M5-M6.5}$  & $0.108-0.122$ & $0.82-0.99$ & $33.1\pm3.4$ & $18.9\pm2.0$ \vspace{2pt} \\ 
  LSPM J1423+1318B & $15.6\pm0.10$ & $14.4\pm0.11$ & $1.20\pm0.15$ & $\rm{M5-M6.5}$  & $0.099-0.109$ & & & \\ 
\\
  LSPM J1735+2634A & $15.3-15.5$ & $13.6-13.8$ & $1.52-1.85$ & $\rm{M7-L3}$  & $0.077-0.086$ & $0.85-0.96$ & $10-12$ & $2.8-3.6$\vspace{2pt}  \\ 
  LSPM J1735+2634B & $16.1-16.5$ & $14.4-14.7$ & $1.69-2.11$ & $\rm{M8-L4}$  & $0.066-0.082$ & & & \\ 
\\
  LSPM J1809+2128A & $15.6\pm0.14$ & $14.4\pm0.15$ & $1.17\pm0.21$ & $\rm{M5-M6}$  & $0.124-0.109$ & $0.82-1.03$ & $41.8\pm4.4$ & $46.8\pm5.0$ \vspace{2pt}\\ 
  LSPM J1809+2128B & $16.0\pm0.11$ & $14.8\pm0.12$ & $1.21\pm0.16$ & $\rm{M5-M6}$  & $0.101-0.112$ & & &  \\ 
 \hline
\end{tabular}
\end{table*}

\noindent\textbf{LSPM J1735+2634}
This system contains a possible brown dwarf companion, and is located at only 10-12pc. The faintness of the companion leads to difficulties in estimating resolved photometry of this system, so only upper and lower limits are noted for many of this system's properties.

The companion has a lower mass limit of 0.066$\rm{M_{\odot}}$ and thus may be a brown dwarf; the allowed mass ranges are detailed in figure \ref{FIG:Mass_model}. Changing seeing conditions and the very red colour of the system gave a more clearly resolved image in the z' filter; in this image the secondary appears very elongated approximately (not exactly) East-West. This elongation suggests that it is possible that the secondary is an unresolved ($\sim0.04''$) brown dwarf binary. This very close pair would have a very short period and can thus be rapidly confirmed with followup observations of the elongation direction.

This system is an excellent target for resolved infrared photometry or spectroscopy to confirm the nature of the secondary. With an expected orbital period of approximately 15-30 years it is also well suited for astrometric followups to measure a dynamical mass.

\noindent\textbf{LSPM J1809+2128} 
A 1.12'' binary; proper motion 0.19''/year, estimated distance $41.8\pm4.4$pc. With an orbital radius of $46.8\pm5.0$AU this estimated M5/M5.5 system is one of the very few known VLM binaries wider than 30AU \citep{Siegler_2005, Phan05}.

\begin{figure}
  \centering
	\subfigure{\resizebox{1.5in}{!}{\includegraphics{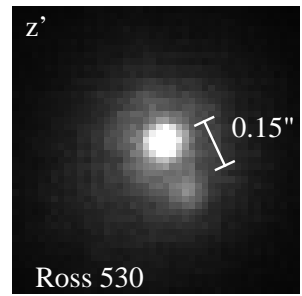}}}
  \caption{The companion to Ross 530. 5\% selection from 900 frames in SDSS z'. The 0.15'' binary is very clearly resolved; the diffraction limit of the telescope in z' is 0.09''.}
  \label{FIG:Ross_530}
\end{figure}

\subsection{A companion to Ross 530}
During the observations we also detected a companion to Ross 530, one of the SDSS standard stars (figure \ref{FIG:Ross_530}). The system is very clearly resolved at only 0.15'' separation.

Ross 530 is known to be a metal-poor spectroscopic binary \citep{Latham_2002}, but to our knowledge this is the first resolved image of this system. From this single observation it is unclear if the single-lined spectroscopic binary has been resolved or if a new companion has been detected. 

\section{Discussion}
\label{SEC:Discussion}
\subsection{The binary frequency of M5.5-M8.0 stars in this survey}
We have detected 5 new binaries in a 32 star sample, giving a raw binary fraction of $16^{+8}_{-4}\%$. However, this survey is based on a magnitude limit that assumes all the targets are single stars. Unresolved binaries in the LSPM survey appear to be brighter and thus closer than single stars for a specific colour, as there are two luminous bodies. The observed binary fraction is thus biased as a result of leakage of binaries into the sample from further distances.

We can compensate for the distance bias by comparing the volume containing the single stars in our sample to the larger space which would contain any binaries. Lacking a useful constraint on the contrast ratios of low mass binaries we assume a range of possible distributions - from all binaries being equal magnitude systems to a flat distribution of contrast ratios \citep{Burgasser_2003}. Including the probability distribution of the raw binary fraction and the range of contrast ratio distributions yields a distance bias corrected binary fraction of $7^{+7}_{-3}\%$.

The derived binary statistics are very similar to the sample of 36 M6.0-M7.5 M-dwarfs described in \citet{Siegler_2005}, who obtain a distance-corrected binary fraction of $7^{+4}_{-2}$.
 
\subsection{Contrast Ratios}
\label{SEC:Conrats}
We did not detect any companions at SDSS i' contrast ratios $>$1 magnitude, although our survey is sensitive to up to 5 magnitude differences (figure \ref{FIG:Sens_curves}), well into the brown dwarf regime. 

\begin{figure}
  \centering
  \resizebox{\columnwidth}{!}
   {
	\includegraphics{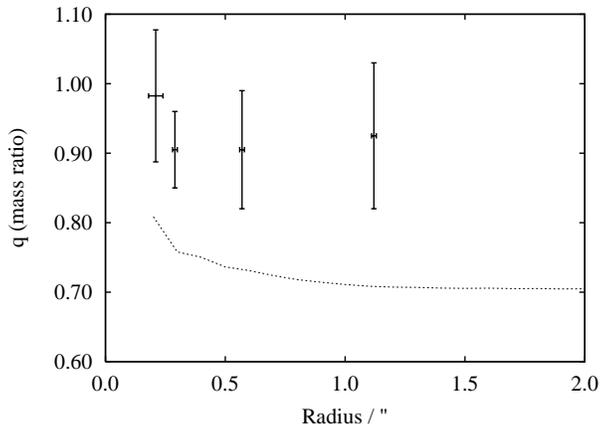}
   }
   \caption{The binaries' mass ratios; an example curve shows the minimum detectable binary mass ratio for a typical star in our survey, an M6.5 primary at approximately 20pc. The contrast ratios shown in figure \ref{FIG:Sens_curves} were used to give a minimum detectable companion mass from the 5.0 Gyr models given in figure \ref{FIG:Mass_model}.}
   \label{FIG:Sens_q}
\end{figure}

The new binaries are all close to equal mass (figure \ref{FIG:Sens_q}), in common with other VLM binary surveys which are sensitive to faint companions (eg. \citealt{Close_2003, Siegler_2005}). 

\subsection{The distribution of orbital radii}

\begin{figure}
  \centering
  \resizebox{\columnwidth}{!}
   { 
	\includegraphics{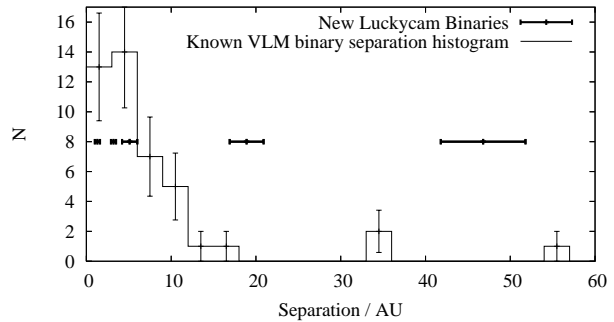}
   }
   \caption{The $1\sigma$ range of orbital radii for each detected binary, compared with a histogram of the previously known sample collated in \citet{Siegler_2005} (and the 33 AU VLM binary described in \citealt{Phan05}). Poisson $1\sigma$ error bars are shown for the histogram - and illustrate the necessity for increased sample sizes. For reasons of clarity, the 200AU system described in \citet{Luhman04} is not displayed.}
   \label{FIG:Sep_hist}
\end{figure}

The new binaries' orbital radii are shown in figure \ref{FIG:Sep_hist}, compared to the distribution of known VLM binary systems with primaries later than M6 (from \citealt{Siegler_2005}). Three of the five systems fall within 1-5AU, the most common radius for known VLM binaries (the surveys are incomplete at very small radii). However, one binary (LSPM J1809+2128) is one of only very few \citep{Siegler_2005, Phan05} known VLM binaries wider than 25AU. It is important to enlarge the sample of VLM binaries to ascertain how common these wide systems are, as well as to constrain the fraction of higher order multiple systems.

\section{Conclusions}
\label{SEC:Concs}
LuckyCam's very low time overheads allowed a 32 VLM star sample to be imaged at high angular resolution in only 5 hours on a 2.5m telescope. Uniquely, this survey was performed in the visible, and thus complements the near-infrared surveys of \citealt{Bouy_2003, Burgasser_2003, Close_2003, Gizis_2003, Delfosse2004} and \citealt{Siegler_2005}. Lucky Imaging shows great promise for surveys of this type, offering a unique ground-based faint-guide-star visible-light imaging capability. 

We have detected five new VLM binaries in a 32 star sample, giving a raw binary fraction of $16^{+8}_{-4}\%$ and a distance-bias corrected fraction of $7^{+7}_{-3}\%$. Primaries were M5.5 to M8; most secondaries were only slightly redder than the primaries. However, one newly found system is a possible M-dwarf brown-dwarf binary. The distribution of orbital radii is in broad agreement with previous results, with a peak at 1-5AU, but one newly detected binary is very wide, at $46.8\pm5.0$AU. No systems with a high contrast ratio were detected, even though the survey is sensitive well into the brown dwarf regime.

\section*{Acknowledgements}
The authors would like to particularly thank Graham Cox at the NOT for help during the observations, as well as many other helpful NOT staff members. We would also like to thank Cathie Clarke, John Baldwin and Peter Warner for many helpful discussions, and we are grateful for the useful comments of an anonymous referee. NML acknowledges support from the UK Particle Physics and Astronomy Research Council (PPARC). This research has made use of the SIMBAD database, operated at CDS, Strasbourg, France. We also made use of NASA's Astrophysics Data System Bibliographic Services.
\bibliographystyle{mn2e}
\bibliography{binaries}

\end{document}